# Tear of Lipid Membranes by Nanoparticles

*Mériem Er-Rafik\*, Khalid Ferji\*, Jerôme Combet, Olivier Sandre, Sébastien Lecommandoux, Marc Schmutz, Jean-François Le Meins & Carlos M. Marques\**

Dr. M. Er-Rafik, Prof. J. Combet, Dr. M. Schmutz, Dr. C. M. Marques
Institut Charles Sadron, Université de Strasbourg, CNRS-UPR 22, 23 rue du Loess, 67034 Strasbourg Cedex 02, France
E-mail: marques@unistra.fr, khalid.ferji@univ-lorraine.fr

Prof. K. Ferji, Dr. O. Sandre, Prof. S. Lecommandoux, Prof J.-F. Le Meins
University of Bordeaux, LCPO UMR 5629, 16 avenue Pey Berland, F-33600 Pessac, France
CNRS, LCPO UMR 5629, 16 avenue Pey Berland, F-33600 Pessac, France

Dr. M. Er-Rafik
Technical University of Denmark, DTU Nanolab, National Centre for Nano Fabrication and Characterization, Building 307, 2800 Kgs Lyngby, Denmark
merafik@dtu.dk

Dr. K. Ferji
Ecole Nationale Supérieure des Industries Chimiques, Laboratoire de Chimie Physique Macromoléculaire, 1 rue Grandville BP20451, 54000 NANCY France

Keywords: lipid membranes, nanoparticles, silica particles, cryo-electron microscopy, nano-toxicity

**Abstract.** Health concerns associated with the advent of nanotechnologies have risen sharply when it was found that particles of nanoscopic dimensions reach the cell lumina. Plasma and organelle lipid membranes, which are exposed to both the incoming and the engulfed nanoparticles, are the primary targets of possible disruptions. However, reported adhesion, invagination and embedment of nanoparticles (NPs) do not compromise the membrane integrity, precluding direct bilayer damage as a mechanism for toxicity. Here it is shown that a lipid membrane can be torn by small enough nanoparticles, thus unveiling mechanisms for how lipid membrane can be compromised by tearing from nanoparticles. Surprisingly, visualization by cryo transmission electron microscopy (cryo-TEM) of liposomes exposed to nanoparticles revealed also that liposomal laceration is prevented by particle abundance. Membrane destruction results thus from a subtle particle-membrane interplay that is here elucidated. This brings into a firmer molecular basis the theorized mechanisms of nanoparticle effects on lipid bilayers and paves the way for a better assessment of nanoparticle toxicity.

Frequent contact between particles of nanoscopic dimensions and the lipid membranes of living organisms is an ineluctable consequence of the rising abundance of nanoparticles in modern technological applications.[1,2,3] The fate of such contacts between particles and membranes, whether accidental or purposeful, is arguably a key determinant to evaluate levels of nano-toxicity or the efficiency of strategies for nano-therapy. But a decade of scrutiny of the transformations induced by nanoparticles on model lipid membranes has only reported mild consequences for the bilayers.[4-8] For hydrophilic particles, moderate affinity for the membrane surface leads to attachment to the outer membrane leaflet, in a partially wrapped state.[9] For larger adhesion energies wrapping becomes complete and the particle is invaginated from the outer to the inner side of the exposed membrane.[10] In a liposome, this leads to a final structure where the particle has crossed the membrane barrier into the liposome lumen in a wrapped – thus protected – state.[5] Hydrophobic particles can either stay embedded in the inner bilayer or, when properly decorated, cross from one side to the other,[7] a process referred to as translocation. In all cases, the membrane is left unperturbed.
In the quest for unveiling the conditions under which the interaction between nanoparticles and membranes might lead to serious membrane disruption, we investigated colloidal suspensions containing liposomes of DOPC, a zwitterionic lipid forming liquid-phase bilayers of 5 nm thickness,[11] and $SiO_2$ nanoparticles of comparable radii. At such small particle dimensions it is foreseen that, if wrapping occurs, membrane deformations will be pushed well beyond their elastic deformation limit, thus increasing the likeness of irreversible membrane damage.

**Figure 1** displays cryo-TEM images of SiO$_2$ nanoparticles, of DOPC liposomes and of their mixtures at a ratio of five nanoparticles per liposome, along with a schematic representation of the typical features observed as one increases the time lapse after the nanoparticles have been added to the liposome suspension. As Figure 1.A shows, the SiO$_2$ nanoparticles are round shaped, albeit not perfectly spherical. Their average diameter is 12.2 ± 1.7 nm, as determined from the mean and the standard deviation of the size distribution, also displayed in the figure, sampled from over one hundred particles. Comparable dimensions where also obtained from small angle X-ray scattering (SAXS) experiments of particle suspensions, see Supporting Figure S1. A typical image of the DOPC liposomes, shown in Figure 1.B, confirms that our preparation method yields a majority of unilamellar structures, with a small fraction of multilamellar liposomes. The liposome size distribution, also presented in the figure, provides values for average size and mean deviations at 125±34 nm. Size distributions from measurements of the bilayer thickness, displayed in Figure 1.Q, yields 4.7±0.5 nm, in agreement with literature values.[11,12] Comparable conclusions can be drawn from small angle neutron and X-ray scattering, SANS and SAXS, as well as from static and dynamic light scattering, SLS and DLS, as displayed in the Figures S1 and S7. Under our conditions, the attractions between the nanoparticles and the liposomes result in a fast attachment. Figure 1.C. shows an image of the system 20 s after the NPs have been added to the liposome suspension. As the image and its corresponding schematic representation in Figure 1.D. show, all the NPs are found to be in contact with the liposome bilayers, albeit with little development yet of wrapping, a feature that is associated with the majority of NP-membrane contacts in the samples imaged later, one minute after addition of the NPs. Typical images of such samples are shown in Figures 1.E and 1.F with the corresponding schematic representations displayed in Figures 1.G and 1.H. At this stage, most of the NPs have been partially wrapped by the lipid bilayer, a typical situation consisting in membrane wrapping of about a half of the NP surface, which can be perceived from both top and side views of the membrane deformations around the particles. Note that top views alone cannot discriminate between partial wrapping and particle engulfment by the liposome. In our case, we confirmed that engulfment does not occur by imaging tilted samples under the electron beam. Partial wrapping results also in the presence of bridges between the membranes of different liposomes, a situation depicted in Figure 1.H. Partial wrapping is not however the final equilibrium state resulting from NP-membrane interactions. As Figures 1.I and 1.J show, exposure of the membranes to the NPs over a period of 30 minutes results in many events where the membrane is torn by the nanoparticles. The sequence of tear formation can be deduced by careful inspection of the micrographs, it is summarized in Figures 1.K and 1.L. In a first step, a partially wrapped NP lacerates the membrane in a localized incision, allowing further wrapping to occur by the membrane to which the NP stayed attached. For identification we named this step as "tear and wrapping". Continuation of this process results in a complete wrapping of the NPs and the consequent detachment of a fully wrapped particle from the membrane. We call this step "wrapping and detach". Observation of the membrane and NP system over longer period of times confirms the destruction potential of these interactions as unwrapped NPs can no longer be observed in the micrographs. After two months, all the particles are protected by a phospholipid bilayer, the dominant spatial distribution exhibiting many bunches of NPs, see Figs 1.M and 1.O. This long-time, slow evolution eventually leads after four months to a suspension of fully wrapped NPs well dispersed in the suspension as shown in Figs 1.N and 1.P. At this point measurements of the thickness of the membranes wrapping the NPs can be performed, to reveal that the thickness of the attached bilayers is comparable to that of free liposomes, the difference of 1 nm between the mean values in the histograms of Figures 1.Q and 1.R being due to the missing head region in contact with the silica surface that cannot be captured by the electronic contrast.

Analysis of cryo-TEM images and the scattering experiments described in the Supporting Information provides thus a clear scenario for the sequence of events and for the key factors leading to membrane tear and to the ensuing membrane destruction. As a prime factor, a strong attraction between the NPs and the lipid bilayer is at play in this system. While in pure NP colloidal solutions the particles are mostly seen as single particles, well dispersed with minor degree of aggregation, for all the suspensions containing NPs and membranes the images show, at any stage of the mixture, that all NPs are adhered to a bilayer. The major direct consequence of this affinity is the deformation of the membrane as it progressively wraps around the NPs. Membrane deformations are also known to mediate attractive interactions between the adhered NPs[13,14]. Here, such attractions are clearly evidenced by the numerous particle-particle contacts seen in the micrographs, often resulting in pearl-like arrangements of NPs. As the membrane progressively wraps around the NPs a point is reached where the membrane is torn at the triple contact line between the free membrane, the adhered membrane and the free particle surface. Wrapping proceeds then further until a fully covered NP detaches from the bilayer. The long-term consequence is complete wrapping of all particles.

Our results show that less than 20 seconds are required for all NPs to encounter and bind to the membranes, a short time compatible with a diffusion limited mechanism for the adhesion between NPs and liposomes. Indeed, under our conditions, the diffusion time of a NP of 10 nm diameter over a typical distance between liposomes (~ 500 nm) is smaller than one second. Also, the first images at 20 seconds already exhibit particle organization at the membrane surface, showing that the development of the membrane deformation fields and the consequent membrane mediated interactions between the particles occur on fast time scales. However the steps that led to further particle wrapping and eventually to membrane rupture are much slower. Further wrapping requires of order of one minute, rupture is only seen after tens of minutes and several months are necessary to completely enwreathe all the NPs and destroy the membranes. Such large time scales can only be understood by long-lived phenomena, such as the collective lipid rearrangement that would be required for significant membrane thinning.

A combination of factors contributes to the membrane rupture. Most strikingly, the radii of the NPs are comparable to the bilayer thickness. Wrapping leads thus to high bending, a deformation state that imposes strong stresses in the two leaflets of the bilayer. The middle plane of the hydrophobic region of the outer leaflet, located at a distance of 3.5 nm from the NP surface is stretched by 13% with respect to the middle bilayer plane at 2.5 nm. For uniform stretching, lysis of the DOPC bilayer is known to occur above membrane tensions of the order of a few mN/m, at deformations between 5% and 10% [11]. For the upper bound of 10%, this simple argument would predict that NPs with radii below 7.5 nm can induce large enough stresses to rupture the bilayer. For our case, the combined effect of the adhesion forces and the curvature generates thus a wrapped bilayer state that is close to the rupture point. Maximal gradients of bilayer properties are obviously located at the triple contact line, where imposed stresses overcome membrane cohesive forces when, according to our estimation, the angle between the membrane and the particle surface exceeds 90º, the point of half wrapping. That a minimal angle is required to rupture the membrane is confirmed in a striking manner by experiments performed with a high particle to liposome ratio r, see **Figure 2**, and Figure S2.

For all the ratios explored here, r=50, r=140 and r=280 no membrane rupture was detected, despite the strong coverage of the membranes by the particles, see also Figure S4. Simple geometric considerations show that a particle of 5 nm radius requires 0.5% of the area of a 50 nm radius liposome to be half wrapped. Since it is unlikely that liposomes possess more than 5 or 10 % of excess available area, angles of the triple line as large as 90º cannot be achieved if more than 10 or 20 NPs are attached to the membrane. In our case, amongst all the ratios tested, experiments (not shown) with r=8 was the largest still leading to membrane tearing.

As a summary, we reported here evidence for lipid bilayer tearing by $SiO_2$ nanoparticles of 12 nm diameter and identified strong adhesion, high curvature and a large degree of wrapping as the key factors that compromise membrane integrity. We expect these key results to hold across a variety of more complex membrane systems and NPs suspensions, and in particular to provide useful guidelines for the evaluation of NP toxicity in biological systems.

**Experimental Section**

*Chemicals and materials.* DOPC, (1, 2-dioleoyl-sn-glycero-3-phosphocholine) was purchased from Avanti Polar Lipids Inc. and has been used to form LUVs. Ludox® SM colloidal silica nanoparticles were purchased from Sigma-Aldrich.

*Sample preparation.* Vesicles were prepared by the sonication method, using the protocol described in the Supporting Information. In order to disperse the NPs, the solution was sonicated in the bath during 15 min before added to DOPC vesicles. The SM30 NPs initial concentration obtaining isolated NPs have been determinate by cryo-TEM and SAXS analyses.

*Cryo-Transmission Electron Microscopy (cryo-TEM).* Images were taken at room temperature from vitrified samples. Sample vitrification was carried out in a homemade vitrification system. The samples were observed under low dose conditions in a Tecnai G2 microscope (FEI) at 200 kV. Images were acquired using an Eagle slow scan CCD camera (FEI). In order to check inner liposome content some images were observed at different tilting angles by tilting the cryo-holder. The size distribution of the NPs, the lipid vesicle diameter and the membrane thickness were measured with the Analysis software (SIS-Olympus, Münster, Germany). The size of NPs and lipid vesicles were determined with the circular shape predefined by the software. The membrane thickness of the lipid vesicle and the one surrounded the NPs were determined by a horizontal density profile. Membrane thickness was measured for the free liposomes between the darkest values of the radial grayscale profiles, which corresponds to the distance between the lipid heads. For the bilayers attached to the particles the head region in contact with the silica is not visible, the distance was measured between the external heads (darkest external value of the grayscale profile) and the end of the internal hydrophobic leaflet (last light greyscale level of the profile). Differences between both methods are of order of 1 nm, the thickness of the head region. For each experiment, the diameters and the membrane thicknesses of more than 60 liposomes and 170 NPs were measured and averaged.

**Acknowledgements**
We thank support from the Agence Nationale de la Recherche under grant ANR-12-BS08-0018. M. Er-Rafik and K. Ferji contributed equally to this work.

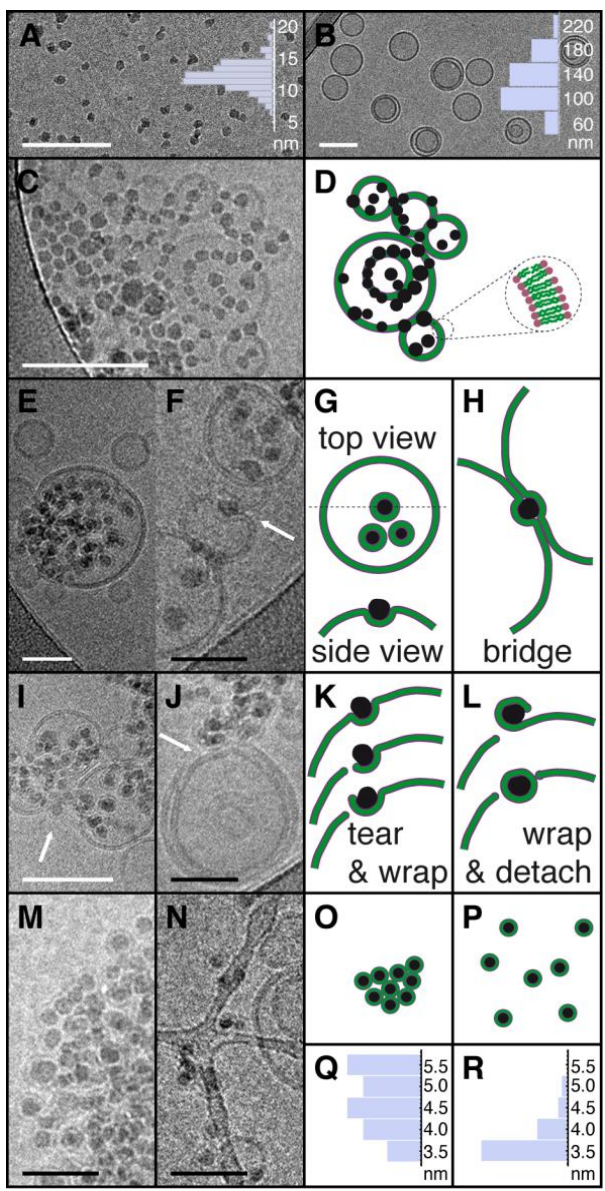

**Figure 1.** ((Cryo-TEM images of SiO$_2$ nanoparticles (SM30 Ludox®) and DOPC liposomes at nanoparticles per liposome ratio r = 5 with corresponding schema highlighting the relevant interaction aspects. **A**. Image of the SM30 particles in a dilute brine solution of 10 mM KCl and the associated diameter distribution with average diameter 12.2 ± 1.7 nm. **B**. Image of the DOPC liposomes in 10 mM KCl. Liposome diameter distribution is also represented with average diameter 125±34 nm. **C**. Image of a NP/liposome mixture, 20 seconds after the nanoparticles have been added to the liposome solution. **D.** Schematic representation of a chosen area in the C image. The particles are shown in black, while the membranes are displayed as a hydrophobic tail layer in green and a hydrophilic head region in purple. All the particles are in contact with the lipid bilayer. **E**. and **F.** Images of the NP/liposome suspension one minute after mixing. **G**. Top and side views of the NPs partially wrapped by the membrane of the larger liposome in E. **H.** A schematic illustration of the interaction topology in image F where one particle bridges the bilayers from two neighbor liposomes. **I.** and **J.** Images after 30 minutes of NP/liposome contact. **K**. Schematic representation of the tearing event pointed by the arrow in figure I, caused by a NP which is not yet completely wrapped by the lipid bilayer. **L.** Schematics of a NP completely wrapped following the tearing event shown in image J. **M.** and **N.** The NP/membranes systems respectively two and four months after mixing. **O.** The scheme of image M shows NPs completely wrapped by the bilayer membrane. Here, the NPs are not yet dispersed. **P.** Fully wrapped and dispersed NPs as shown in image N. **Q.** Thickness distribution of the bilayers from liposomes not exposed to NPs, showing an average thickness of 4.7±0.5 nm measured from head to head. **R.** Thickness distribution of the bilayers completely wrapping the nanoparticles from four months old samples, with an average thickness of 3.7±0.3 nm measured from the head position far from the silica surface to the tail-head interface on the silica surface. White scale bar 100 nm, black scale bar 50 nm.))

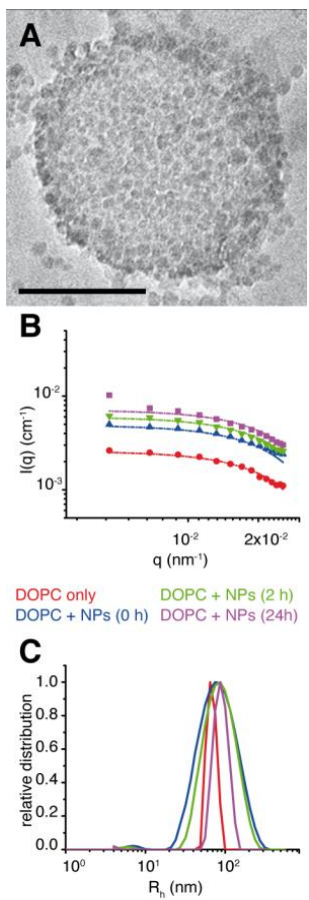

**Figure 2.** ((Large nanoparticle per liposome ratios r stabilize the membranes. **A.** Cryo-TEM image of a NP/liposome mixture with r=50, one minute after the nanoparticles have been added to the liposome solution, scale bar 100 nm. All particles adhere to the liposomes, without any visible membrane disruption. **B.** Light scattering spectrum from a mixture of r=280 at different stages after mixing and **C.** the corresponding size distribution from DLS measurement of the same systems.))

**Table of content entry**

Lipid bilayers are not compromised by exposure to nanoparticles, they display instead phenomena such as engulfment or embedment. Are then nanoparticles only toxic through indirect biochemical mechanisms like protein imbalance or oxidation? In this work it is shown that *if nanoparticles are small enough they do lead to direct physical damage of lipid bilayers* by tearing.

**Bilayer-nanoparticle interactions**

M. Er-Rafik*, K. Ferji*, J. Combet, O. Sandre, S. Lecommandoux, M. Schmutz, J.-F. Le Meins and C. M. Marques*

**Tear of Lipid Membranes by Nanoparticles**

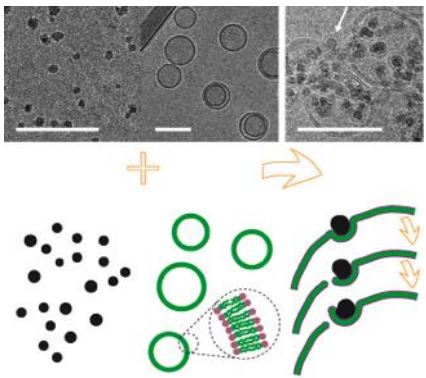

# Supporting Information

**Tear of Lipid Membranes by Nanoparticles**

*Mériem Er-Rafik\*, Khalid Ferji\*, Jerôme Combet, Olivier Sandre, Sébastien Lecommandoux, Marc Schmutz, Jean-François Le Meins & Carlos M. Marques\**

**Supporting Figures**

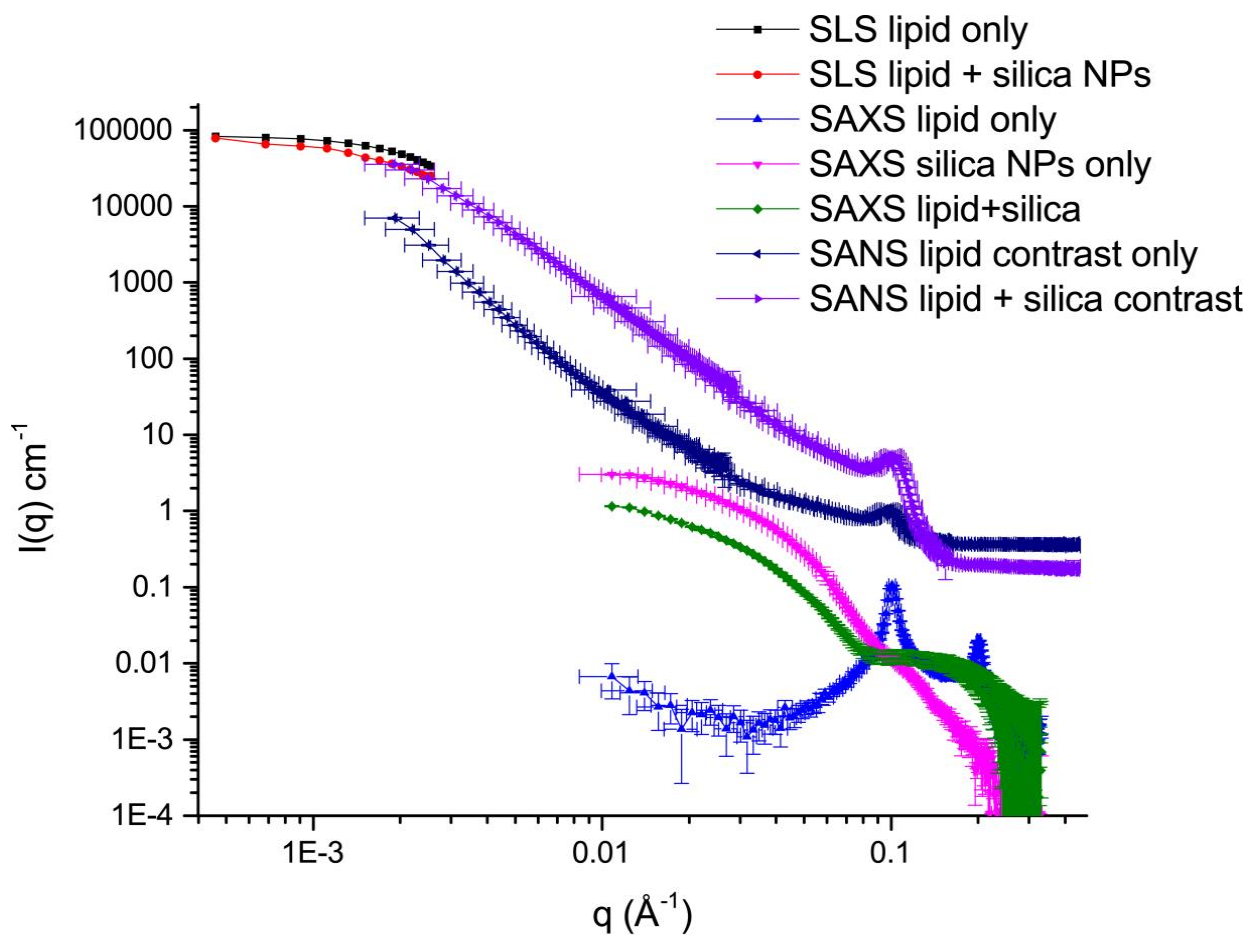

**Figure S1.** Static light scattering (SLS), small angle X-ray and neutron scattering experiments (SAXS and SANS) of liposomes, NPs and of mixtures of liposomes and NPs.

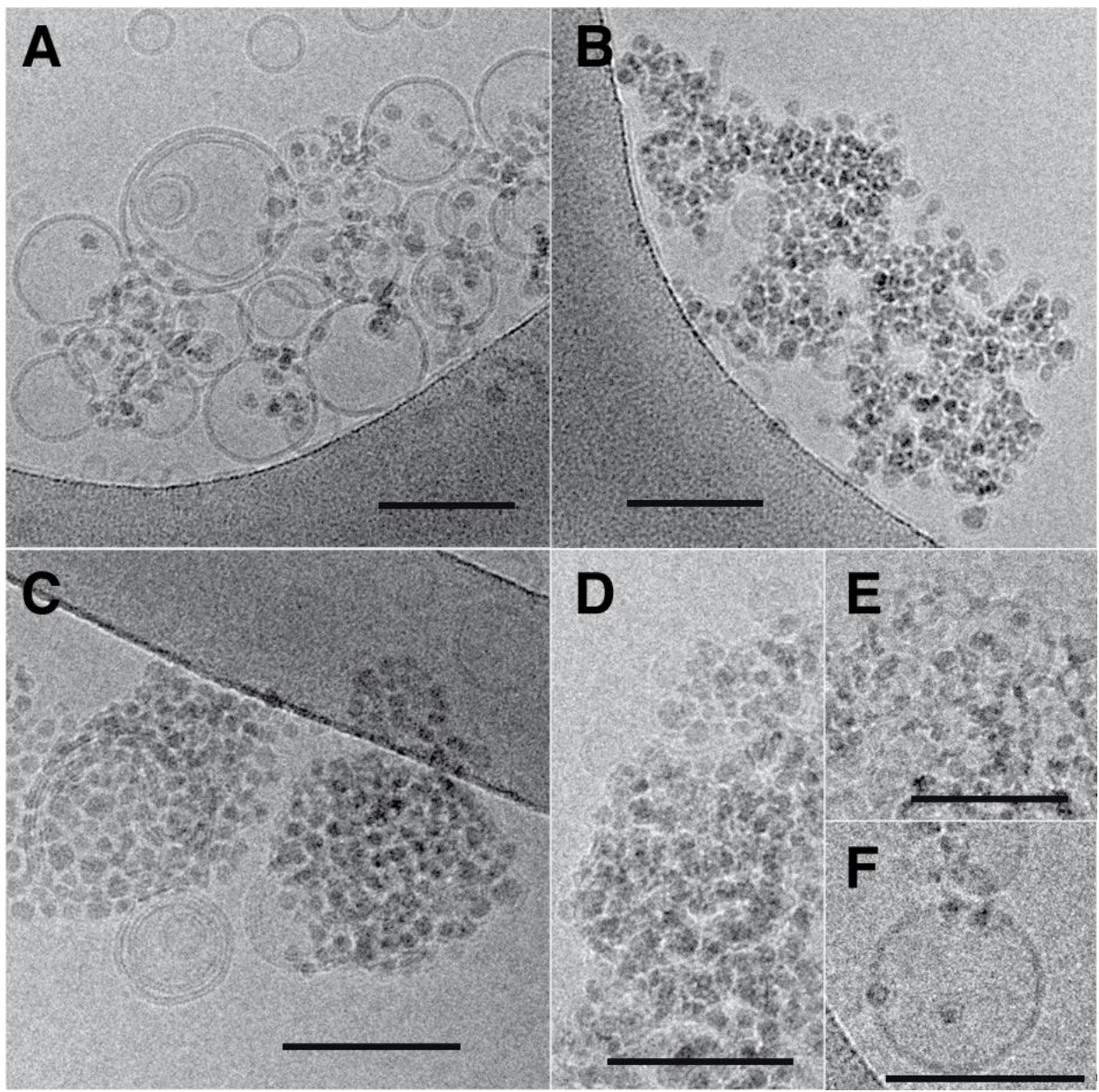

**Figure S2.** Cryo-TEM images of SiO$_2$ nanoparticles (SM30 Ludox®) and DOPC liposomes at nanoparticles per liposome ratio r = 5, at different pH. **A**. Image of the SM30 particles in a brine solution of 10 mM KCl at pH 7, one minute after the nanoparticles have been added to the liposome solution. **B**. Similar conditions after 30 minutes. **C**. Image of the SM30 particles in a brine solution of 10 mM KCl at pH 10, one minute after the nanoparticles have been added to the liposome solution. **D, E** and **F**. Similar conditions after 30 minutes.

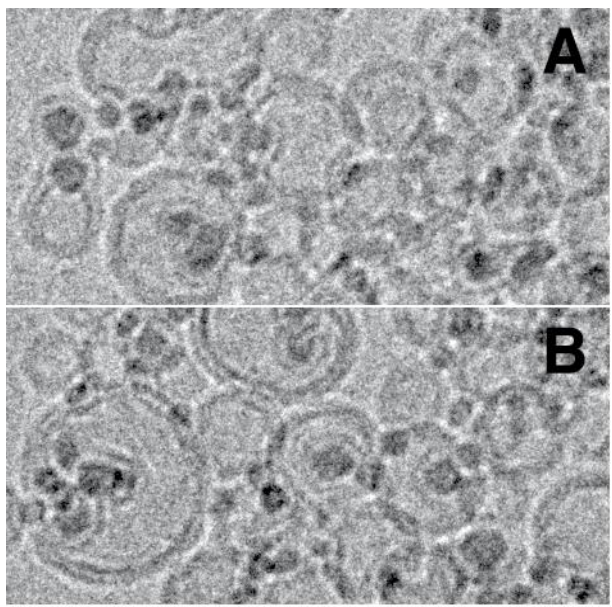

**Figure S3.** Cryo-TEM images of SiO$_2$ nanoparticles (SM30 Ludox®) and DOPC liposomes at nanoparticles per liposome ratio r = 5, at pH 5.5. **A**. Image of the SM30 particles in a solution without added salt one minute after the nanoparticles have been added to the liposome solution. **B**. Similar conditions after 30 minutes.

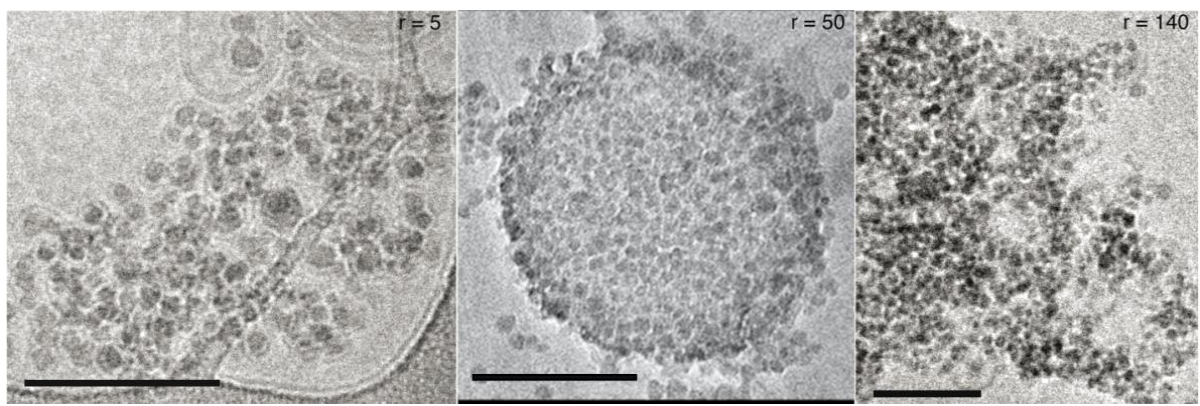

**Figure S4.** Effect of the number ratio r between the SiO$_2$ nanoparticles (SM30 Ludox®) and DOPC liposomes, in a brine solution of 10 mM KCl at pH 5.5, one minute after the nanoparticles have been added to the liposome solution. Cryo-TEM images of the SM30 particles and DOPC liposomes at **A**. r=5, **B**. r=50 and **C**. r=140.

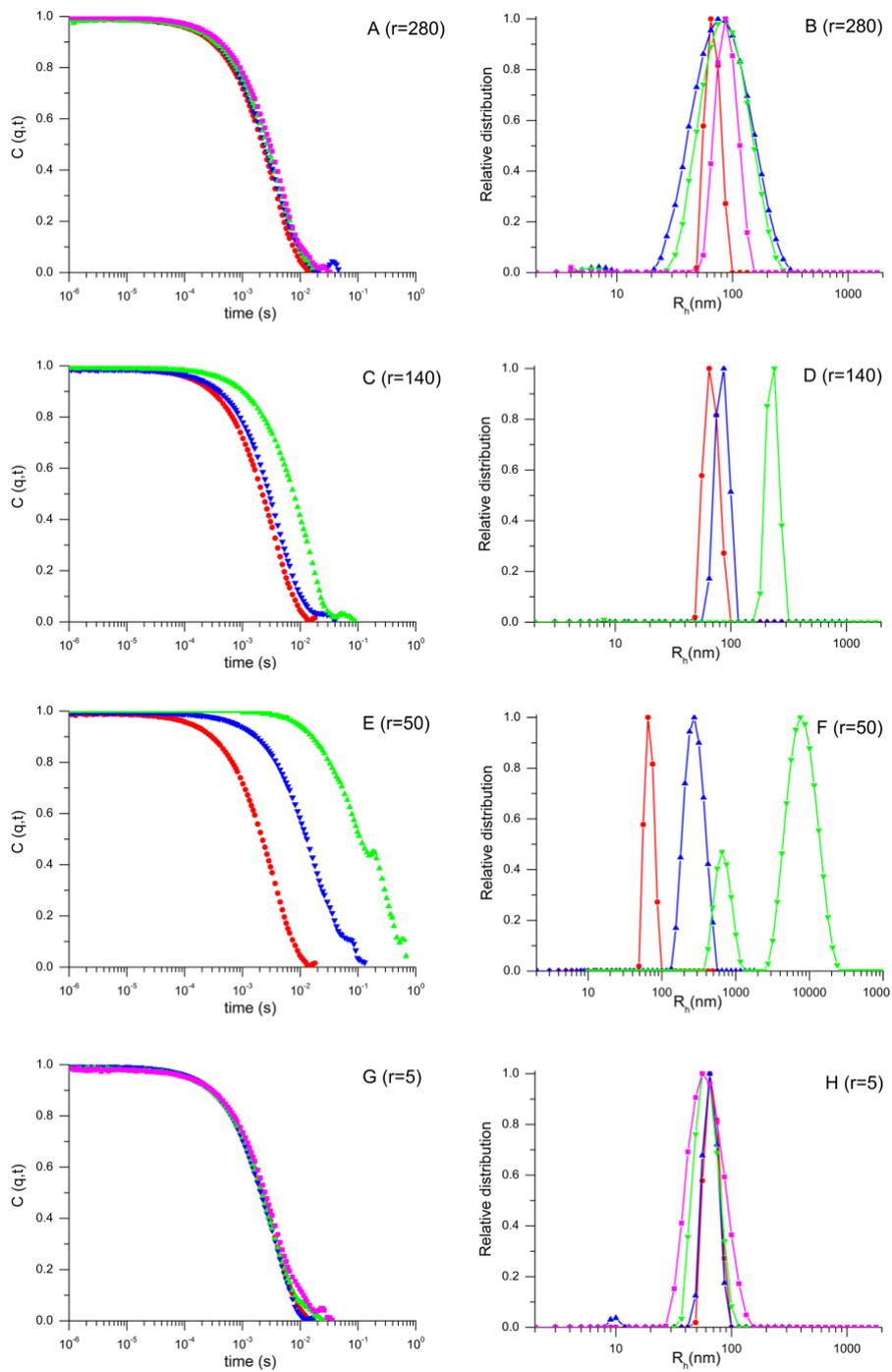

**Figure S5.** Left panels display autocorrelation functions C(q,t) as a function of time and right panels the corresponding relative distribution of relaxation times recorded at a scattering wave vector q = 6.84 × 10$^{-3}$ nm$^{-1}$ (θ = 30°) for free liposomes (●) and for different mixtures of nanoparticles per liposome ratios (A, B) r=280, (C, D) r=140, (E, F) r=50, (G, H) r=5. Analysis realized at different time interval; (▲) immediately after preparation, (▼) after 2h, (■) after 24h. Dashed lines are given as guides to the eye.

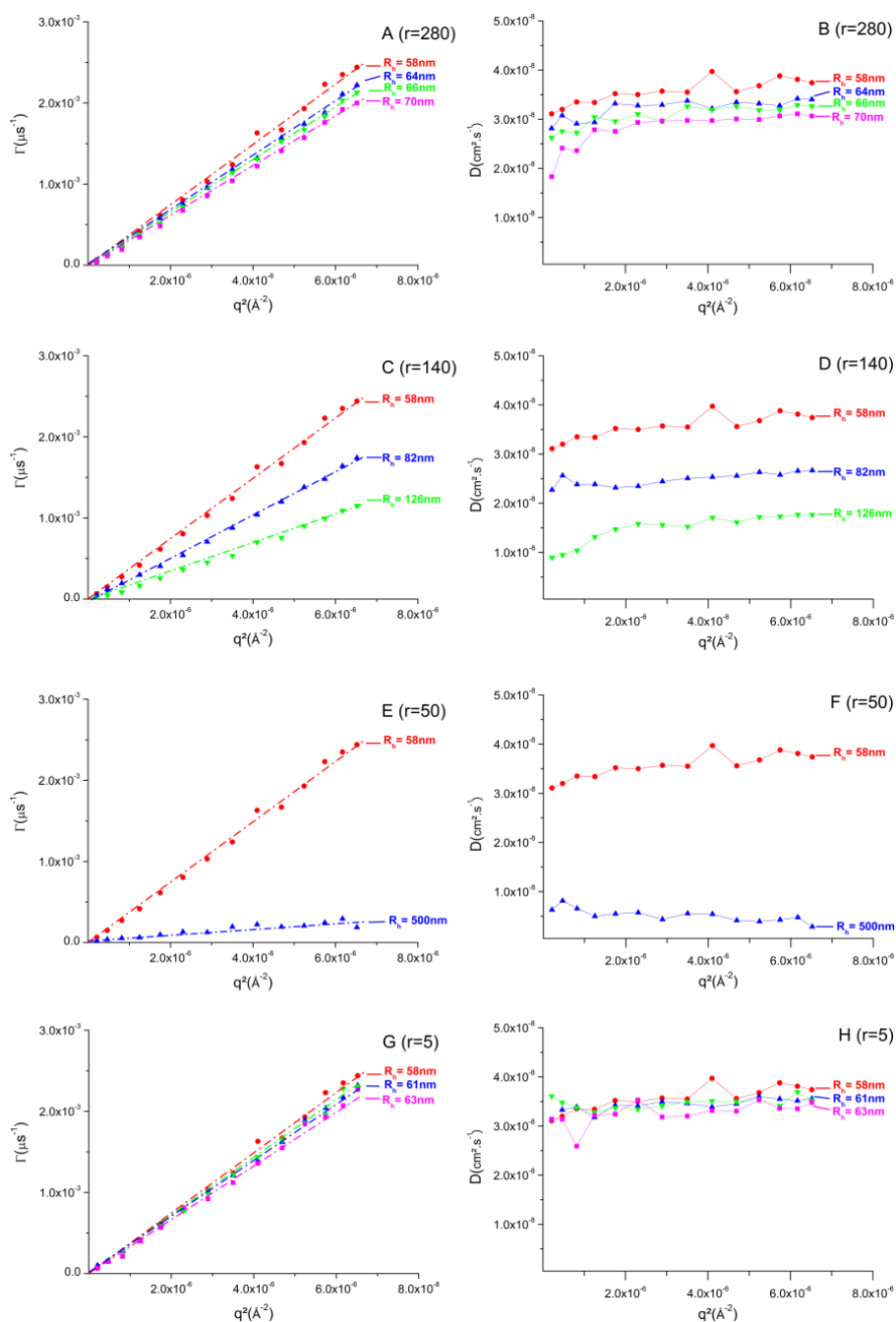

**Figure S6.** Left panels display the q² dependence with relaxation rate (Γ) and right panels the corresponding q² dependence with diffusion coefficient at scattering vector (q) ranging between 4.6 x 10-3 nm-1 and 2.55 x 10-2 nm-1 for free liposomes (●) and for different mixtures of nanoparticles per liposome ratios (A, B) r=280, (C, D) r=140, (E, F) r=50, (G, H) r=5. Analysis realized at different time interval; (▲) immediately after preparation, (▼) after 2h, (■) after 24h. Dashed lines are given as guides to the eye.

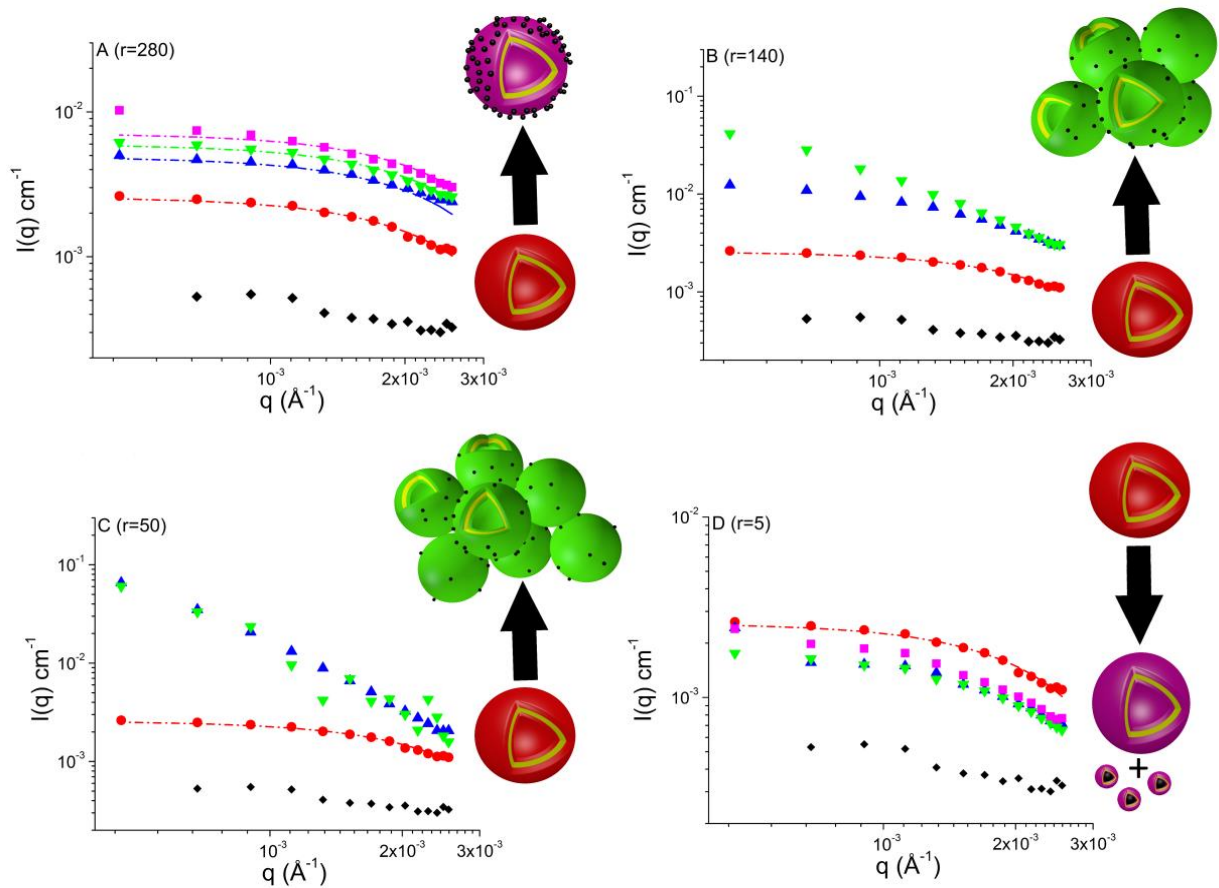

**Figure S7.** Total light scattered intensity (I) as a function of the scattering wave vector q at 20°C for free nanoparticles (◆), free liposome (●) and for different mixtures of nanoparticles per liposome ratios; (A) r=280, (B) r=140, (C) r=50, (D) r=5. Analysis realized at different time interval; (▲) immediately after preparation, (▼) after 2h, (■) after 24h. Dashed lines are theoretical expressions of light scattered intensity obtained using Eq.4 (described below in the supporting information, parameters used are listed in table-S1). The schematic representations illustrate the resulting objects after mixing liposomes with $SiO_2$ nanoparticles.

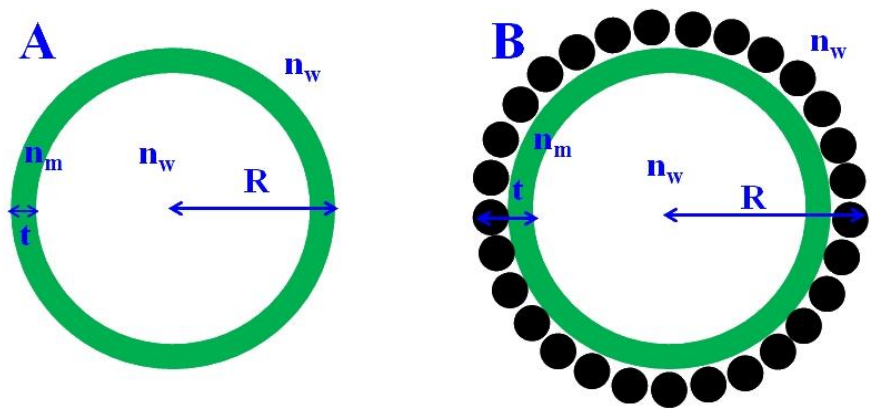

**Figure S8.** Schematic representation of model used in theoretical expression (Eq- S4) of light scattering intensity for free liposomes in the absence (A) and in the presence (B) of $SiO_2$ nanoparticles.

## Supporting Methods

### 1. Sample preparation

#### 1.1 Preparation of DOPC LUVs

a- Cryo-TEM characterization. A solution of DOPC in chloroform/MeOH was evaporated in a rotary evaporator to obtain a thin film. This film was hydrated at room temperature with deionized water (resistivity higher than 18 Ω) containing 10 mM of KCl and homogenized by vigorous stirring to yield 10 mg.mL$^{-1}$ concentrated dispersions of vesicles. Suspensions of liposomes without added KCl where equally prepared. The resulting coarse dispersion was submitted to sonication (Vibracell 75042, 15 mm titanium probe, 16 min, 50% pulse). The resulting transparent dispersion was heated at 40 °C for 1hour in order to anneal defects that could have been produced by sonication.

b- Light scattering measurement. Liposomes were prepared by the film rehydration process followed by extrusion through a polycarbonate membrane. Briefly, a solution (100 mg.mL$^{-1}$) of DOPC in chloroform/MeOH (3v:1v) was prepared, then a thin film of lipid was formed by evaporation of solvents under reduce pressure. The thin film was then hydrated in aqueous solution of 10 mM of KCl under gentle agitation during 2 hours. Afterwards, LUVs (1 mg.mL$^{-1}$) were obtained by extrusion (21 times) through a polycarbonate filter with pore sizes of 100 nm at room temperature.

#### 1.2 Preparation of DOPC/SiO2 nanoparticles mixtures

a- Cryo-TEM characterization. Microscope images from mixtures with nanoparticle per liposome ratio r=5 where prepared from aqueous dispersions (10 mM of KCl) of SiO$_2$ nanoparticles (3 mg.mL$^{-1}$) mixed with liposomal suspensions of DOPC (10 mg.mL$^{-1}$) giving a final concentration of 8.33 mg.mL$^{-1}$ of DOPC liposomes. Other ratios r where prepared from the same initial concentrations of nanoparticles and liposomes, with the proportions of NP dispersions and liposome suspensions required to achieved the target ratios.

b- Light scattering measurements. In order to prepare mixtures with four nanoparticle per liposome ratios (r), r= 5, 50, 140 and 280, aqueous dispersions (10 mM of KCl) of SiO$_2$ nanoparticles (diameter of 10 nm) with appropriate concentrations were added separately to the suspension of liposomes (0.12 g.L$^{-1}$) described above. The final liposome concentration in mixtures was kept constant (0.06 g.L$^{-1}$) for all ratios. Samples were analyzed immediately after preparation and at 2 and 24 hours.

### 2. Cryo-Transmission Electron Microscopy (cryo-TEM).

Images were obtained a vitrified sample, at room temperature. The vitrification of the samples was carried out in a homemade vitrification system. The chamber was held at 22 °C and the relative humidity at 80%. A 5 µl drop of the sample was deposited onto a lacey carbon film covered grid (Ted Pella) rendered hydrophilic using an ELMO glow discharge unit (Cordouan Technologies). The grid is automatically blotted to form a thin film which is plunged in liquid ethane hold at -190°C by liquid nitrogen. In that way a vitrified film is obtained in which the native structure of the vesicles is preserved. The grid was mounted onto a cryo holder (Gatan 626) and observed under low dose conditions in a Tecnai G2 microscope (FEI) at 200 kV. Images were acquired using an Eagle slow scan CCD camera (FEI).

## 3. Dynamic and static light scattering (SLS and DLS)

Experiments were performed with a multibit, multitau, full digital ALV/CG-3 correlator in combination with a Spectra-Physics laser (22 mW HeNe linear polarized laser operating at wavelength of λ= 632.8 nm) and a thermostat bath controller (20 °C). Measurements were made at angles θ ranging between 20° and 150° with corresponding scattering vectors q ranging from q=4.6 x $10^{-2}$ $nm^{-1}$ to 2.55 x $10^{-1}$ $nm^{-1}$. The autocorrelation functions $g_1(t)$ were analyzed in terms of relaxation time (τ) distribution (Eq-S1) according to the REPES method (see examples at θ=30°, Fig. S5). Hydrodynamic radius ($R_H$) was estimated using the Stokes-Einstein relation (Eq-S2), where $D_0$ is diffusion coefficient obtained from the relaxation rate (Γ) (see Fig. S6), $K_B$ is the Boltzmann constant (1.38 $10^{-23}$ $J.K^{-1}$), T is experimental temperature (20°C) and $\eta_s$ is the viscosity of the solvent (water, 1 Pa.s).

$$g_1(t) = \int A_{(\tau)} \exp\left(\frac{-t}{\tau}\right) d\tau \qquad \text{Eq-S1}$$

$$R_H = \frac{K_B T}{6\pi \eta_s D_0} \qquad \text{Eq-S2}$$

For each angle θ, rayleigh ratio ($R_\theta$) of samples, recorded by SLS measurements was obtained using (Eq–S3) as shown in fig. S7. Where $I_{solution}$, $I_{water}$, and $I_{toluene}$ are the intensities scattered by the solution, water (solvent) and toluene (reference), respectively. $n_{water}$ = 1.33 and $n_{toluene}$ = 1.496 are the refractive indexes, and $R_{toluene}$ = 1.35 ×$10^{-5}$ $cm^{-1}$ is the Rayleigh ratio of toluene for a wavelength λ = 632.8 nm.

$$R_\theta = \frac{I_{solution}(\theta) - I_{water}(\theta)}{I_{toluene}(\theta)} \times \left(\frac{n_{water}}{n_{toluene}}\right)^2 \times R_{toluene} \qquad \text{Eq-S3}$$

The large optical contrast between liposome ($n_{DOPC} = 1.52$), $SiO_2$ nanoparticles ($n_{SiO_2} = 1.45$) and water ($n_{H_2O} = 1.33$) is favorable to study the interaction process of nanoparticles with liposomes. To this aim, the Rayleigh ratio $R_\theta$ was fitted by a modified Rayleigh-Gans-Debye approximation (RGDA) of the total scattered intensity $I$ for a coated sphere using (Eq-S4) [1,2,3], where $f = 1 - \frac{t}{R}$, with R the radius of vesicle and t the thickness of lipid bilayer as schematically depicted in Fig. S8-A; $x = R \times q$, with q the scattering vector; $j_1(x)$ is the first-order spherical Bessel function; $m_1 = \frac{n_m}{n_w}$ with $n_w$ and $n_m$ the refractive indexes of aqueous medium (1.33) and membrane, respectively. For free liposomes, $n_m$ is estimated to be of 1.52 while in case of the nanoparticles/liposomes mixture, the membrane refractive index $n_m$ was estimated from the refractive indexes of the $SiO_2$ nanoparticles ($n_{NPS}$=1.45) and of the lipids ($n_{lipid}$ 1.52) using the mass fraction of each component ($y_{NPs}$ of silica nanoparticle and lipid ($y_{NPs}$)) (Eq-S5).

$$I = \left((m_1 - 1)\left(\frac{3 \times j_1(x)}{x} - f^3 \times \frac{3 \times j_1(f \times x)}{f \times x}\right)\right)^2 \qquad \text{Eq-S4}$$

$$n_m = y_{lipid} \times n_{Lipid} + y_{NPs} \times n_{NPs} \qquad \text{Eq-S5}$$

### 3.1 DOPC/SiO2 interaction:

In order to study the interaction of DOPC liposomes ($R_h$= 58 nm) with small $SiO_2$ nanoparticles ($R_h$= 5 nm), four aqueous $SiO_2$ nanoparticle solutions were mixed separately with a diluted aqueous liposome solution with a constant concentration of 0.12 g.$L_{-1}$ as described above. Four mixtures with different nanoparticle per liposome ratio r were then prepared; high ratio (r=280), intermediate ratios (r= 50, 140) and low ratio (r=5).

Fig. S7 displays the scattered intensity as a function of the scattering vector for liposome solutions without nanoparticles and with nanoparticles at four ratios. In the case of the liposome solution, the scattered intensity was well fitted by theoretical expression described above (Eq. S4) using parameters listed in table-S1. Fig. S7 shows also that immediately after mixing nanoparticles with liposomes the light scattering intensity becomes, for all solutions, different from the sum of the single components (liposomes and nanoparticles), indicating the presence of rapid and strong interactions between $SiO_2$ nanoparticles and liposomes. In addition, compared to the scattering intensity of liposomes, a shift of values and deviation of shapes are observed after adding nanoparticles depending on the nanoparticles per liposome ratio used, indicating the presence of different interaction processes as discussed below.

### 3.2 Case-1: High ratio of nanoparticles per liposome (r=280):

Fig.S5-A shows the relaxation functions C(q,t) as a function of time for liposomes before and after mixing with nanoparticles at r = 280. A small shift of the relaxation function is observed in comparison to the free liposomes - see fig. S5-A - indicating a small change of the diffusion coefficient of the scattering objects as depicted also in figs. S6-A,B. The diffusion coefficients of the liposome-nanoparticle mixtures are smaller than that of the liposomes, indicating a small increase of hydrodynamic radius $R_h$, 58 ± 3 nm in the absence and 66.6 ± 3 nm in the presence of nanoparticles.

Fig.S7-A shows the light scattering intensity as a function of the scattering vector q for liposomes before and after adding nanoparticles at r = 280. The shape of the scattering function is preserved indicating a conservation of the vesicular morphology of liposomes upon addition of the nanoparticles, and therefore shows that there is no cluster formation. In addition, a significant rising of scattered intensity is observed for the resulting objects comparing to free liposomes, indicating strong interaction between liposomes and nanoparticles as described above. Landfester *et.al.* [2] have observed similar phenomenon after adding $SiO_2$ nanoparticles having higher hydrodynamic radii (14, 25, 36 and 57 nm) to PDMS-b-PMOXA polymersomes. Authors have associated such increase of light scattering to increasing values of inner polymersome refractive index due to nanoparticles incorporation into the interior of polymersomes by endocytosis mechanism [2]. This mechanism is unlikely in our case given that nanoparticle crossing of the membrane requires membrane consumption leading to vesicle shrinkage [2], while figs S6-A,B show that in our case the size of resulting objects is almost similar to pure liposomes. The increase of the scattered intensity observed in fig. S7-A can be associated to the adsorption and saturation of liposomes surface by nanoparticles leading to hybrid liposomes as schematically described in fig. S7A. By considering such scenario (see fig. S8-B), the scattering intensity of the resulting objects

could be fitted well with the theoretical expression (Eq-S4) using a higher thickness (12.5 ± 0.15 nm) and modified refractive index of hybrid membrane (1.466) as listed in table-S1.

**3.3 Case-2: intermediate ratios of nanoparticles per liposome (r= 50 and 140):**

Figs.S7-B,C show the light scattered intensity as a function of the scattering vector q for liposomes before and after adding nanoparticles at r = 140 and r = 50. In both cases, an increase of scattered intensity is observed after mixing liposomes with nanoparticles, indicating a strong interaction of liposomes with nanoparticles as described above. For these ratios the shape of the scattering intensity are significantly modified by the addition of nanoparticles, the resulting objects are modified comparing to the free liposomes, indicating change of vesicular morphology and therefore does not allow fitting light scattering by the theoretical expression (Eq-4) as it previously made at r= 280. Such deviations are associated to cluster formation as schematically presented in figs. S7-B,C, which is corroborated by the rise of the resulting objects sizes depicted in figs. S6-C,D and figs. S6-E,F. For both ratios, a white solid is formed after 24h mixing. Cluster formation could be attributed to the presence of bridges formed by nanoparticles between the membranes of different liposomes given the strong attraction between DOPC and $SiO_2$ nanoparticles, and to the low coverage of liposome surfaces by nanoparticles given the lower r used here.

**3.3 Case-3: low ratio of nanoparticles per liposome (r= 5):**

Fig.S5-G shows the relaxation function C(q,t) as a function of time for liposomes before and after mixing with nanoparticles at low nanoparticle per liposome ratio (r=5). No shift of the relaxation function is noticed after adding the nanoparticles, indicating that the resulting size of the objects remain similar to the ones of bare liposomes as it is demonstrated by the diffusion coefficients D(q) shown in figs.S6-G,H. However, unlike observations made above at high ratios (r=280, 140 and 50), fig. S7-D shows that when a small amount of nanoparticles is added to the liposomes (r=5), the light scattered by the suspension decreases immediately, from its initial value after mixing. Such phenomena is consistent with the partial destruction of liposomes exposed to a small number of nanoparticles, which causes a reduction of the liposome concentration in the suspension. Although the cryo-TEM images (see fig.1) clearly show that liposome destruction also leads to nanoparticles wrapped by a lipid bilayer, this new species do not contribute enough to the signal to be detected. It is also important to stress that, despite the decreasing values scattered light intensity, the shapes of the scattered functions do not change, indicating that the vesicles remaining in solution are preserved, without the formation of clusters.

Light scattering experiments thus confirm the scenario presented in the main text of the paper. At high nanoparticles per liposome ratio, hybrid nanoparticles/liposomes objects are produced by adhesion of nanoparticles on liposomes surface. At intermediate ratios, cluster formation is observed. However, when low amount of nanoparticles is mixed with liposomes, partial liposome destruction occurs and wrapped nanoparticles by lipid bilayer are produced.

Table-1 Parameters used in theoretical expression (Eq. 4) of light scattering intensity for free liposomes and mixture at nanoparticles per liposome ratio of 280.

|  |  | $n_{wa}$ | $n_{mb}$ | $R_c$ | $T_d$ |
|---|---|---|---|---|---|
| Free liposomes | | 1.33 | 1.52 | 58nm | 5nm |
| r=280 | Immediately | 1.33 | 1.466 | 60nm | 11nm |

| | after mixing | | | | |
|---|---|---|---|---|---|
| | 2 hours | 1.33 | 1.466 | 60nm | 12.5nm |
| | 24 hours | 1.33 | 1.466 | 60nm | 14nm |

a Refractive index of water. b Refractive index of membrane. c Hydrodynamic radius of vesicle. d Thickness of vesicle membrane.

## 4. Small angle X-ray and neutron scattering (SAXS-SANS)

Small angle X-ray scattering experiments have been performed with a Rigaku microfocus rotating anode generator (Micromax-007 HF) operating at 40kV and 30 mA (Cu-K radiation, $\lambda$= 1.54 Å). The X-ray beam was monochromatized and focused with a confocal Max-Flux Optics developed by Osmics, Inc. together with a three pinholes collimation system. Scattered intensity was measured with a 2D multiwire detector located at 0.81 m from the sample. This configuration allowed q vectors to be investigated in the range 0.01 Å$^{-1}$< q <0.33 Å$^{-1}$ (q is defined as q=$4\pi/\lambda$sin ($\theta$/2) where $\lambda$ is the wavelength of the incoming beam and $\theta$ is the scattering angle). The suspensions were held in calibrated mica cells of 1 mm thickness. Scattering patterns were corrected according to the usual procedures. Data were radially integrated, corrected for electronic background, detector efficiency, empty cell scattering and transmission. Scattering from the pure solvent was measured separately and removed from the sample measurements. Intensity was converted into absolute scale using calibrated Lupolen as a standard. Measurements have been performed on pure NP and DOPC suspensions as well as on DOPC-NP mixtures.

Small angle neutron scattering experiments have been performed two weeks after sample preparation at the Orphée neutron facility of LLB-CEA Saclay on the PAXY spectrometer with an anisotropic two-dimensional detector. Dispersions of SiO$_2$ NPs (2.8 g.L$^{-1}$) mixed with DOPC liposomes (48 g.L$^{-1}$) at a number ratio r=5 where prepared in two solvent compositions: in pure D$_2$O, of neutron scattering length density $SLD$(D$_2$O)=6.40×10$^{-6}$ Å$^{-2}$, where the neutron scattering signal originates both from silica, $SLD$(SiO$_2$)=3.32×10$^{-6}$ Å$^{-2}$, and from DOPC, $SLD$(DOPC)=0.30×10$^{-6}$ Å$^{-2}$, and in a D$_2$O:H$_2$O mixture (56:44 v/v) matching the neutron scattering length density of silica, in which the signal arises from the phospholipids only. Three beamline configurations were used to cover overlapping scattering vector ($q$) ranges of 1.92×10$^{-3}$ – 2.84×10$^{-2}$, 1.05×10$^{-2}$ – 0.154, and 3.19×10$^{-2}$ – 0.427 Å$^{-1}$, with the following values of sample-to-detector distance $D$ and neutron wavelength $\lambda$: $D$=7 m and $\lambda$=15 Å, $D$=3 m and $\lambda$=6 Å, $D$=1 m and $\lambda$=6 Å. The scattering intensity curves were divided by the transmission factor and subtracted from the incoherent background, before normalizing by the flat signal of a cuvette filled with light water to correct the detector efficiency, yielding the absolute intensity in cm$^{-1}$.

**Nanoparticles suspensions**

NPs have been characterized in very dilute suspensions i.e. in a concentration range where only their form factor contributes to the scattered intensity (the inter-particle structure factor S($q$)=1)). Analysis with a spherical model (homogeneous scattering length density $\rho$=18.65 10$^{10}$ cm$^{-2}$) reveals clear deviations from a pure monodisperse system. A model of polydisperse spheres forming very small clusters (program SASview) has been used to determine the geometrical properties of the NP. The mean diameter of the spherical subunits is 9 nm ($\sigma$=0.25) while clusters contain two elementary spherical particles. This description is in accordance with cryo-TEM measurement that leads to a typical size of the order of 12 nm, in between the average diameter of one spherical subunits (9 nm) and the average length of the NP doublets (18 nm).

**Nanoparticles – DOPC suspensions**

SAXS experiments were performed on the same suspension as measured by cryo-TEM while SANS measurements was done at much higher concentration and shorter delay after preparation (two weeks). After two weeks, the typical form factor of multilamelar vesicles ($q^{-2}$ dependence and characteristic Bragg peaks) is still observable on the SANS curves. Neutron scattering overlaps with SLS at low q (Guinier plateau) and is compatible with the SAXS curve of pure lipids at high q (Bragg peaks at $q$=0.1 and 0.2 Å$^{-1}$). Strong scattering (very similar to that of pure NPs suspensions) is observed by SAXS in the low $q$ region. Furthermore, a large shoulder appears around 0.1Å$^{-1}$ that matches the scattering of pure DOPC suspensions (if we omit Bragg peaks). The most important point is that Bragg peaks (initially present for pure DOPC suspensions) disappear in SAXS measurements (i.e. after four months) while it is still present on SANS measurements after two weeks. The destruction of multi-lamellar vesicles observed after four months evidences the large impact of the NPs on the organisation of the DOPC membranes at long times. SAXS curves cannot be explained on the basis of a single form factor associated to "decorated" NPs (core-shell organisation: NP + adsorbed DOPC layer). More likely, free unilamellar vesicles are still present in the suspension and are responsible for the large shoulder around 0.1 Å$^{-1}$. The system contains a large fraction of unilamellar vesicles (no Bragg peak). Due to the large X-ray scattering length density of the $SiO_2$ NPs with respect to the solvent (but also to the DOPC molecules), their contribution dominates the scattered intensity at low $q$. The adsorbed DOPC layer only contributes at larger $q$. Unfortunately this contribution is hidden by the large shoulder observed above 0.07Å$^{-1}$ that originates from pure and free DOPC vesicles. Therefore, from the scattering experiments, the exact nature of the adsorbed layer (between monolayer or bilayer) cannot be determined.

## Supporting References